\documentclass[conference]{IEEEtran}
\usepackage{cite}
\usepackage{enumitem}
\usepackage{algorithm}
\usepackage{algorithmic}
\usepackage{comment}
\usepackage{amsmath,amssymb,amsfonts}
\usepackage{algorithmic}
\usepackage{graphicx}
\usepackage{textcomp}
\usepackage{xcolor}
\def\BibTeX{{\rm B\kern-.05em{\sc i\kern-.025em b}\kern-.08em
    T\kern-.1667em\lower.7ex\hbox{E}\kern-.125emX}}

\usepackage{mahnigCEC}
\begin{document}

\title{Generating and analyzing small-size datasets to explore physical observables in quantum Ising systems \\
}

\author{\IEEEauthorblockN{1\textsuperscript{st} Rodrigo Carmo Terin}
\IEEEauthorblockA{\textit{University of the Basque Country}\\
\textit{Department of Computer Science and Artificial Intelligence}\\
\textit{Intelligent Systems Group} \\
San Sebastian, Spain \\
rodrigo.carmo@ehu.eus}
}
\maketitle

\begin{abstract}
  We propose a detailed analysis of datasets generated from simulations of two-dimensional quantum spin systems using the quantum Ising model at absolute zero temperature. Our focus is on examining how fundamental physical properties, energy, magnetization, and entanglement entropy, evolve under varying external transverse magnetic fields and system sizes. From the Quantum Toolbox in Python (QuTiP), we simulate systems with 4, 8, and 16 spins arranged in square lattices, generating extensive datasets with 5000 samples per magnetic field value. The Hamiltonian operator incorporates quantum mechanical effects such as superposition and tunneling, challenging classical interpretations of spin states. We compute extended Pauli operators and construct the Hamiltonian to include spin-spin interactions and transverse field terms. Our analysis reveals that as the system size increases, fluctuations in energy and entanglement entropy become more evident, indicating lifted sensitivity to external perturbations and suggesting the onset of quantum phase transitions. Spin-spin correlation functions demonstrate that interactions are predominantly local, but larger systems exhibit more complex and fluctuating correlations. These findings provide valuable insights into the behavior of quantum spin systems and lay the groundwork for future machine learning applications aimed at predicting physical quantities and identifying phase transitions from a quantum perspective.
\end{abstract}



\section{Introduction} \label{sec:intro}

Predicting phase transitions involves understanding complex systems in which small changes in external conditions can lead to abrupt transformations in the properties of the system. Challenges include dealing with non-linear dynamics, critical fluctuations, and the need for accurate datasets to detect these transitions~\cite{Sachdev2011}. The Ising model is an important tool in physics for studying phase transitions, particularly in ferromagnetism. It helps in being aware of how local interactions can lead to collective behavior, acting as an analogy for neurons in machine learning models where local rules determine the overall behavior~\cite{Terin:2024kzc}. The study of phase transitions faces a significant evolution as it moves into the quantum mechanics framework \cite{Heisenberg1925,Dirac1925}; particularly through the famous two-dimensional ($2$D) quantum Ising model \cite{Mbeng:2020awt}. This expansion into quantum mechanics introduces novel paradigms for understanding phase behavior, transitions, and critical phenomena, especially where traditional local order parameters are absent~\cite{Nandkishore2015}.

Furthermore, several other models are used to investigate phase transitions in both classical and quantum viewpoints. For example, the Potts model~\cite{Potts1952} is a generalization of the Ising model and can be used to study phase transitions in systems with more than two possible states per site. It is useful in areas such as magnetism and biology, where interactions can assume multiple discrete states. In the Heisenberg model~\cite{Heisenberg1928}, spins can be oriented in any direction in a three-dimensional space, unlike in the Ising model, where spins are restricted to two directions. This model is fundamental for understanding magnetism in materials where spin-spin interactions are more complex and not restricted to parallel or antiparallel alignments. The Hubbard model~\cite{Hubbard1963} studies electronic systems in networks of atoms or ions, where the competition between the kinetic energy of electrons and their interactions can lead to several physical phenomena such as interaction-induced insulation and superconductivity. The $XY$ model~\cite{MerminWagner1966} considers spins aligned in a plane, capable of freely rotating within it. Such model investigates the Kosterlitz-Thouless phase transition \cite{kosterlitz1973ordering}, which occurs through the decoupling of vortices and antivortices, a topological phase transition without a change in the symmetry of the order parameter. Finally, the percolation model~\cite{Flory1941} scrutinizes the formation of random clusters and their ability to form a large network that covers the entire system. It is applied to both material science problems and in studies of complex networks and epidemiology, focusing on the dynamics of connectivity.

However, the center of attention of our work is on analyzing datasets generated by simulating quantum spin systems with different numbers of spins at zero temperature, examining their energy, magnetization, and entanglement entropy under varying external magnetic fields. This is justified since we aim to prepare the most suitable datasets for our future machine learning works by making predictions of physical quantities and identifying phase transitions from the viewpoint of quantum systems, analogously to what we previously done in the classical perspective~\cite{Terin:2024kzc}. To achieve this, the Hamiltonian operator is fundamental in quantum mechanics to illustrate how the states and interactions of a system's particles dictate its energy~\cite{Sakurai2011}. Thus, the quantum Ising model is our first choice for analyzing phase transitions in the quantum realm. Its theoretical expansion within statistical physics summarizes ferromagnetic behaviors in magnetic particle systems under quantum effects~\cite{Mbeng:2020awt}. Recognized for its multidimensional applicability, this model goes beyond classical constraints, making it indispensable for phase transition studies~\cite{Dutta2010, PhysRevLett.103.207201, PhysRevA.94.022112}.

Represented by the Hamiltonian operator
\begin{equation}
\hat{H} = - J \sum_{\langle i, j \rangle \in L} \hat{\sigma}_{i}^{z} \hat{\sigma}_{j}^{z} - h \sum_{i \in L} \hat{\sigma}_{i}^{x}, \label{eq:quantum_ising}
\end{equation}
where $J$ is the uniform coupling constant between spins, $h$ is the uniform external transverse magnetic field, $\hat{\sigma}_{i}^{x}$ and $\hat{\sigma}_{i}^{z}$ are the Pauli matrices acting on site $i$, and the sum $\langle i, j \rangle$ runs over nearest-neighbor pairs in the lattice $L$. As mentioned earlier, this system introduces quantum superpositions and tunneling effects, challenging classical parallel and antiparallel states through quantum mechanical rules. The inclusion of the Pauli matrix $X$, $\hat{\sigma}_{i}^{x}$, represents quantum state flips, further complicating the system's dynamics at lower temperatures, where quantum phenomena are more evident~\cite{Pauli1927, Hund1927a, Hund1927b}.

Here, we employ quantum mechanical simulations using the Quantum Toolbox in Python (QuTiP)~\cite{Johansson2012QuTiP, Johansson2013QuTiP2} to generate our datasets. Our primary goal is to analyze the energy, magnetization, and entanglement entropy of these systems and to understand how these properties evolve and stabilize as the system size increases.

These datasets provide perceptions into the stability and behavior of the quantum Ising model with different spin configurations. We fix the coupling coefficient $J$ as $1.0$ (in arbitrary energy units), and work with a range of external magnetic field strengths $h$ from $1.0$ to $5.0$ (in the same energy units), at absolute zero temperature, i.e., considering exclusively quantum effects without thermal fluctuations. Each simulation runs for $5000$ samples per value of the external magnetic field, ensuring a good exploration of the state space.

This paper is organized as follows. In Sec.~\ref{sec:intro}, we introduce the background and motivation for studying quantum phase transitions using the two-dimensional quantum Ising model. In Sec.~\ref{sec:DATA}, we describe the data generation process, including the simulation of quantum spin systems with different numbers of spins and the methods used to compute energy, magnetization, and entanglement entropy. In Sec.~\ref{sec:Anal}, we analyze the results, discussing how these physical quantities evolve with varying external magnetic fields and system sizes. Finally, Sec.~\ref{sec:conclusion} concludes the manuscript, outlining our findings and offering potential directions for future research, particularly the application of machine learning algorithms to predict physical quantities and identify phase transitions from a quantum perspective.

\section{Data Generation} \label{sec:DATA}

The core of our dataset creation involves the use of extended Pauli operators and a Hamiltonian formulation to simulate the quantum spin system under varied physical conditions. Our model parameters include grid sizes corresponding to systems with $4$, $8$, and $16$ spins, organized into square lattices of sizes $2 \times 2$, $2 \times 4$, and $4 \times 4$, respectively.

The process begins with the extended pauli(N) function, which establishes the quantum mechanical framework by generating extended Pauli $\hat{X}$ and $\hat{Z}$ matrices tailored to our $N$-spin system. This function builds up Pauli operators for a lattice of spins by creating extended Pauli $\hat{X}$ and $\hat{Z}$ operators for each position on the lattice. Mathematically, for a system with $N$ spins, the extended Pauli operators are given by:
\begin{equation}
\hat{X}_i = I^{\otimes (i-1)} \otimes \hat{\sigma}_x \otimes I^{\otimes (N-i)},
\end{equation}
\begin{equation}
\hat{Z}_i = I^{\otimes (i-1)} \otimes \hat{\sigma}_z \otimes I^{\otimes (N-i)},
\end{equation}
where $I$ is the identity matrix, and $\otimes$ denotes the Kronecker product (tensor product), used to construct the operators for the entire system from individual spin operators. Here, $I^{\otimes (i-1)}$ indicates the identity matrix acting on the first $i-1$ spins, $\hat{\sigma}_x$ is applied on the $i$-th spin, and $I^{\otimes (N-i)}$ represents $I$ performing on the remaining $N - i$ spins.

Next, the ising hamiltonian (N, J, h, precalculated\_ops) function computes the Hamiltonian operator for the $2D$ quantum Ising model, including both the spin-spin interaction terms $\hat{Z}_i \hat{Z}_j$ and the transverse field terms $\hat{X}_i$. The Hamiltonian is formulated as:
\begin{equation}
\hat{H} = -J \sum_{\langle i, j \rangle} \hat{Z}_i \hat{Z}_j - h \sum_{i} \hat{X}_i,
\end{equation}
in which $J$ is the coupling constant between neighboring spins, and $\langle i, j \rangle$ denotes the summation over all nearest-neighbor pairs in a two-dimensional square lattice with periodic boundary conditions. The function identifies the neighbors for each spin and sums the interaction terms accordingly.

It is important to note that in the above-mentioned Hamiltonian, the transverse magnetic field term $- h \sum_{i} \hat{X}_{i}$ is responsible for introducing quantum effects, e.g., superposition and tunneling. The operator $\hat{X}_{i}$ (the Pauli $\hat{\sigma}_{x}$ matrix) acts as a spin-flip operator, enabling each spin to transition between the $|\uparrow\rangle$ and $|\downarrow\rangle$ states.
Mathematically, this introduces the possibility of transitions between spin states, resulting in a superposition of quantum states. Consequently, the system is no longer confined to fixed classical configurations but can explore a variety of states simultaneously, reflecting the probabilistic nature of quantum mechanics \cite{Heisenberg1925,Dirac1925}.

Furthermore, the transverse-field term enables quantum tunneling between different spin configurations. Even at absolute zero temperature, where there is no thermal energy available, spins can ``tunnel" among states because of the quantum fluctuations induced by the transverse field. This is particularly important in quantum systems, as it allows the system to escape from local energy minima and explore the configuration space more thoroughly. By including this term in the Hamiltonian, our simulations capture these essential quantum effects, allowing for a deeper analysis of the system's behavior under different physical conditions and lattice sizes.

Then, our simulation process proceeds with the generation of random quantum states for each spin, using these states in the evolution under the computed Hamiltonian. The calculate entanglement entropy (state, N) function computes the entanglement entropy of a quantum state. The quantum state $|\psi\rangle$ is first represented by its density matrix $\rho = |\psi\rangle \langle \psi|$, which is next partitioned into two subsystems $A$ and $B$ of equal size. The entanglement entropy is computed using the von Neumann entropy $S$ of the reduced density matrix $\rho_A$:
\begin{equation}
S = - \operatorname{Tr}(\rho_A \log \rho_A),
\end{equation}
where $\rho_A = \operatorname{Tr}_B (\rho)$ is attained by tracing out subsystem $B$. Diagonalizing $\rho_A$ yields its eigenvalues $\{\lambda_i\}$, and the entropy is computed as:
\begin{equation}
S = - \sum_i \lambda_i \log \lambda_i,
\end{equation}
providing the quantum entanglement. This approach, widely used in the study of quantum systems, enables the quantification of entanglement in terms of the information shared between two subsystems, thus identifying the behavior of entangled spin configurations~\cite{Nielsen2000quantum}.

The simulate and calculate entanglement (N, J, h\_values, num\_samples) function has the following simulation process:
\begin{enumerate}
    \item The Hamiltonian operator $\hat{H}$ of the quantum Ising model is built up for given $J$ and $h$.
    \item For each sample:
    \begin{enumerate}
        \item A random quantum state $|\psi\rangle$ is generated.
        \item The following quantities are computed:
        \begin{itemize}
            \item The expected energy:
            \begin{equation}
            E = \langle \psi | \hat{H} | \psi \rangle.
            \end{equation}
            \item The average magnetization:
            \begin{equation}
            M = \frac{1}{N} \sum_{i=1}^{N} \langle \psi | \hat{Z}_i | \psi \rangle.
            \end{equation}
            \item The entanglement entropy $S$ of the reduced density matrix $\rho_A$, as previously mentioned.
        \end{itemize}
    \end{enumerate}
    \item Data are collected for different values of $h$ and stored in a pandas DataFrame.
\end{enumerate}

The results, which include these calculated quantities along with the individual spin states for different parameter values, are compiled into a dataset and stored in a CSV file for further analysis.

\section{Analysis of the quantum Ising model with different spin configurations}\label{sec:Anal}

This section provides an analysis of the two-dimensional quantum Ising model system with varying numbers of spins, specifically focusing on $4$, $8$, and $16$ spins. The analysis includes the following quantities. 
\begin{itemize}
\item{Mean Energy vs External Magnetic Field:}
\begin{figure}[h]
    \centering
    \includegraphics[width=0.5\textwidth]{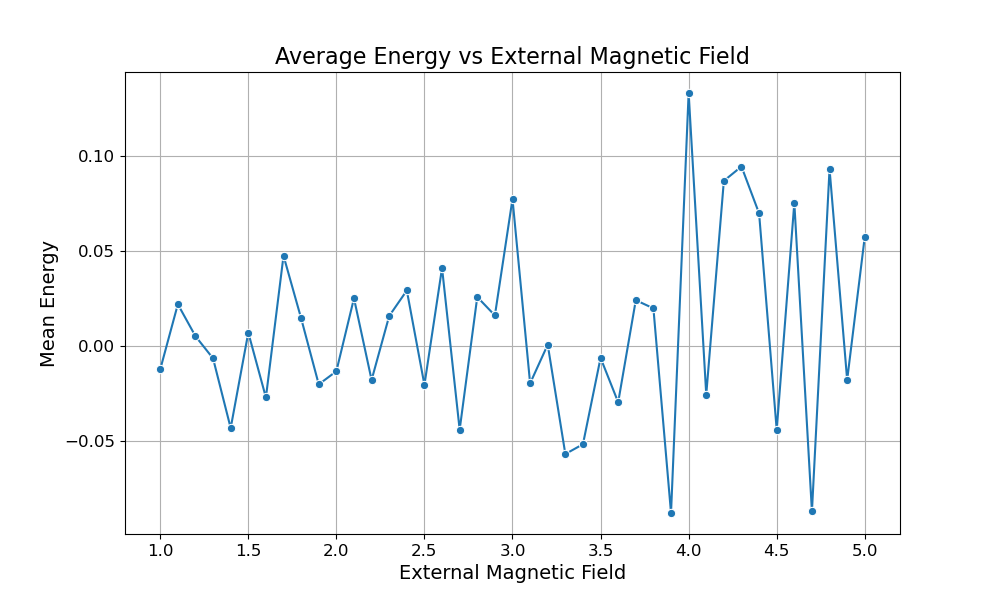}
    \caption{Energy vs External Magnetic Field.}
\end{figure}
The three datasets manifest significant variations in energy in reply to the external magnetic field. With more spins, the differences appear to be more evident, indicating a considerable magnetic response by increasing the number of spins. 

\item{Mean Magnetization vs External Magnetic Field:}
\begin{figure}[h]
    \centering
    \includegraphics[width=0.5\textwidth]{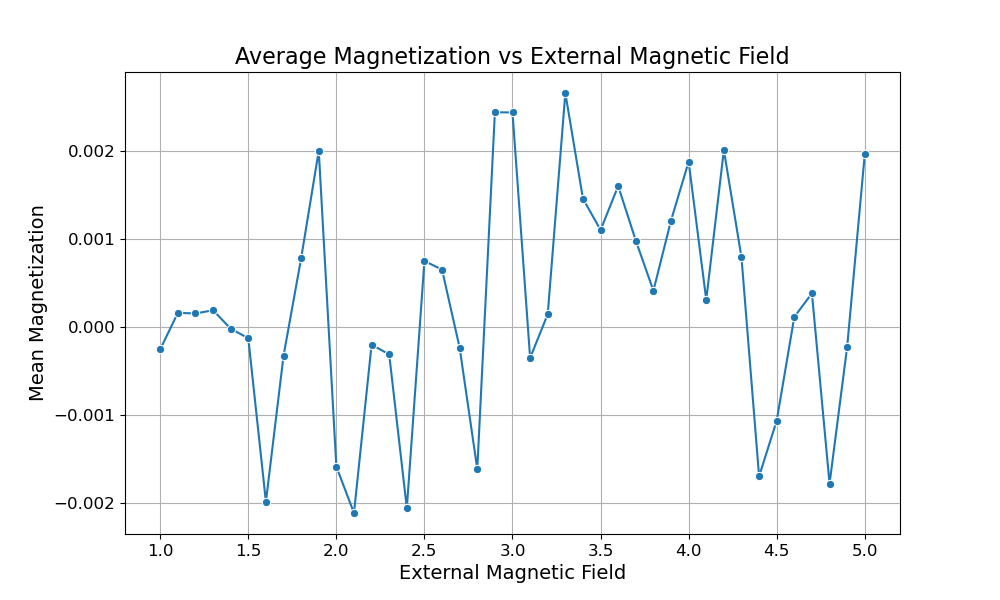}
    \caption{Magnetization vs External Magnetic Field.}
\end{figure}
All datasets show an enlargement in magnetization with the growth in the external magnetic field. The system with $8$ spins exhibits a more linear reply, while the other systems show non-linear behaviors, possibly due to more localized spin interactions. 

\item{Distributions of Energy, Magnetization and Entanglement Entropy:}

\begin{figure}[h]
    \centering
    \begin{minipage}{0.5\textwidth}
        \centering
        \includegraphics[width=\textwidth]{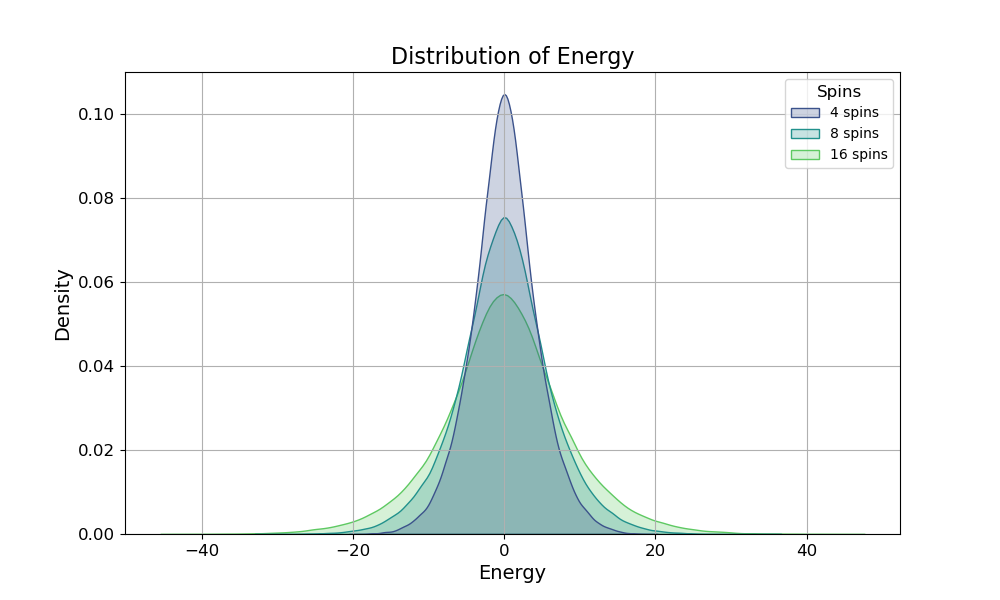}
        \caption{Distribution of Energy.}
        \label{fig:energy_distribution_18_07}
    \end{minipage}\hfill
    \begin{minipage}{0.5\textwidth}
        \centering
        \includegraphics[width=\textwidth]{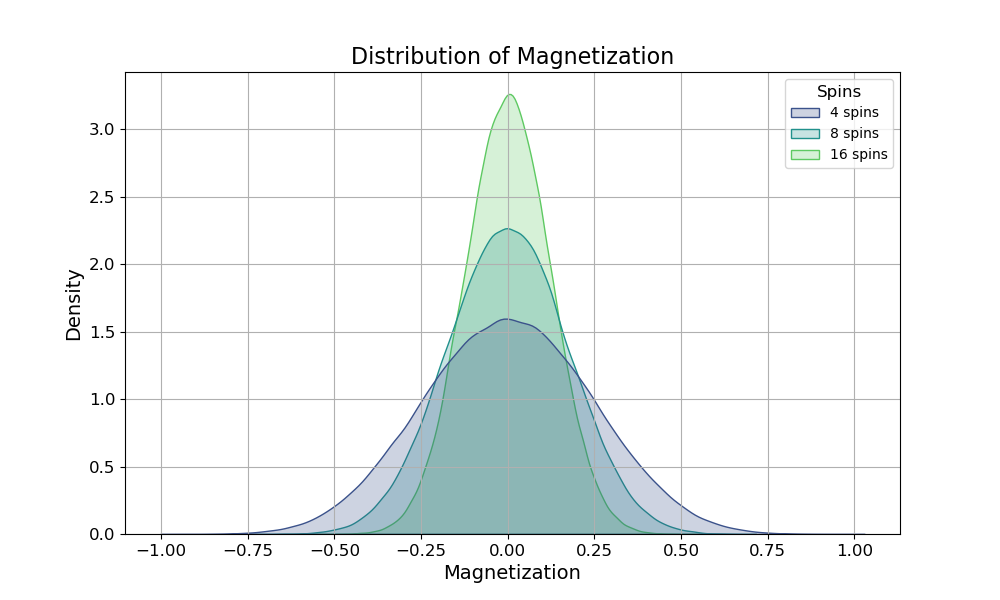}
        \caption{Distribution of Magnetization.}
        \label{fig:magnetization_distribution_18_07}
    \end{minipage}\hfill
    \begin{minipage}{0.5\textwidth}
        \centering
        \includegraphics[width=\textwidth]{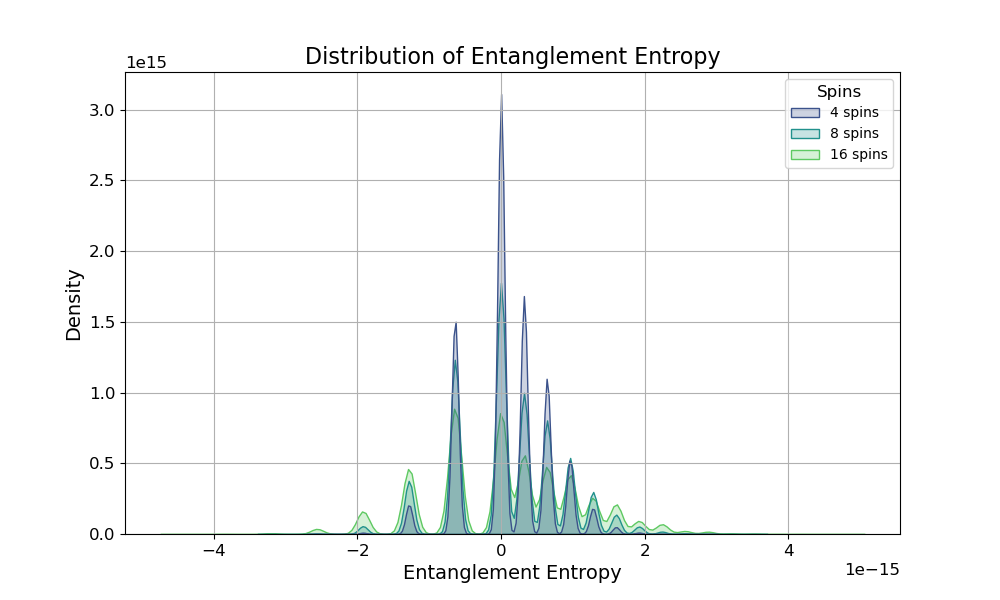}
        \caption{Distribution of Entanglement Entropy.}
        \label{fig:entanglement_entropy_distribution_18_07}
    \end{minipage}
\end{figure}

In Fig \ref{fig:energy_distribution_18_07} the energy distribution for the $4$-spin system shows a wide variability with peaks pointing out to specific stable energy states. In the $8$-spin system, the distribution is better uniform with less variance between energy states, suggesting a greater influence of the increased number of spins. The $16$-spin system exhibits an even smoother distribution with less evident peaks, potentially being a sign a more homogeneous system in terms of energy states. With reference to magnetization distribution displayed in Fig. \ref{fig:magnetization_distribution_18_07}  the $4$-spin system shows considerable variations, reflecting the magnetization's sensitivity to external factors in a smaller system. Growing to $8$ spins, the distribution becomes further centered with less extreme variations, proposing a balance. The $16$-spin system shows an even higher centered distribution with a more defined peak, insinuating a better consistent magnetic alignment over the system. By using Fig. \ref{fig:entanglement_entropy_distribution_18_07} the distribution of entanglement entropy in the $4$-spin system is large with several peaks, pointing to a variety of entanglement states. As the number of spins increases to $8$, the distribution becomes more centralized with fewer high-entropy states. With spins of $16$, it turns into uniform and concentrate around lower values of entropy, possibly reflecting increased coherence in the system. 
\newpage
Next, we continue our analysis by now focusing on examining fluctuations in energy, magnetization, and entanglement entropy and evaluating spin-spin correlations. In the following, we provide interpretations of the results based on the generated plots.

\item{Energy Variance and Stability}

The energy variance plot (Figure \ref{fig:energy_variance_vs_spins}) shows a increase in variance as the number of spins increases from $4$ to $16$. The latter system exhibits much higher energy fluctuations compared to the former- and $8$-spin systems. This suggests that as the number of spins increases, the system becomes sensitive to external magnetic fields, possibly indicating the onset of quantum phase transitions, where larger systems are more susceptible to fluctuations.
\begin{figure}[h]
    \centering
    \includegraphics[width=0.5\textwidth]{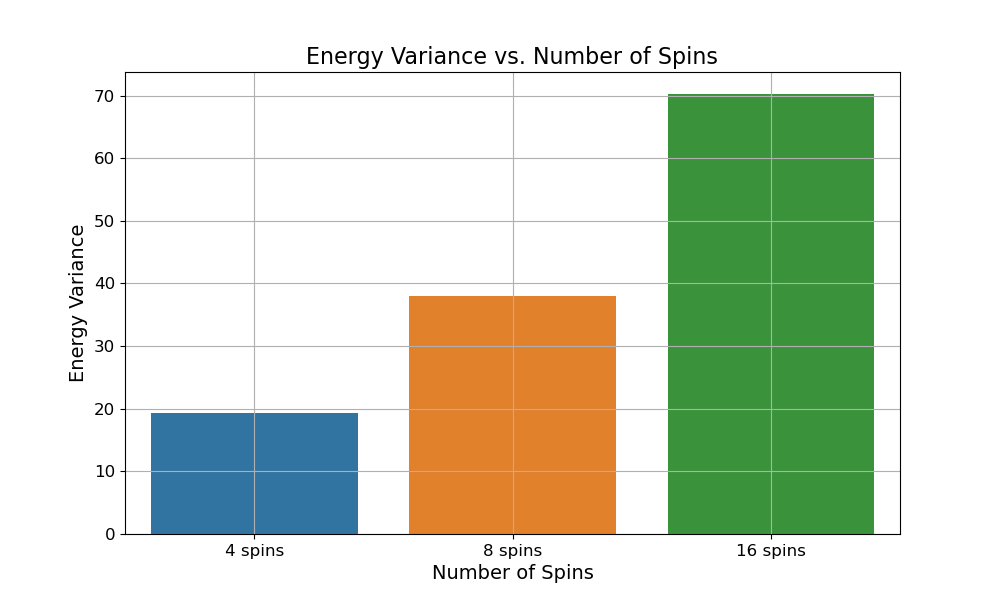}
    \caption{Energy Variance vs. Number of Spins.}
    \label{fig:energy_variance_vs_spins}
\end{figure}

\item{Spin-Spin Correlations}

The spin-spin correlation functions (Figures \ref{fig:spin_spin_correlation_4}, \ref{fig:spin_spin_correlation_8}, and \ref{fig:spin_spin_correlation_16}) and the corresponding table \ref{tab:spin_correlation_all_spins_with_dashes} show how the correlations among spins vary as the distance between them increases for systems of spins $4$, $8$ and $16$. In general, the correlation strength decreases with increasing distance on all systems, indicating short-range spin interactions. For the $4$-spin system (Figure \ref{fig:spin_spin_correlation_4}), the correlation function is weakly negative and exhibits little difference with distance, suggesting limited interaction among neighboring spins and weak overall correlations. In the $8$-spin system (Figure \ref{fig:spin_spin_correlation_8}), the correlation function is further complex, with the highest correlation at the nearest neighbor distance, followed by a steady decline. Small positive correlations persist at intermediate distances, while negative correlations appear at larger distances, indicating patterns of alternating alignment and anti-alignment as the distance increases. The $16$-spin system (Figure \ref{fig:spin_spin_correlation_16}) shows even higher intricate behavior, with oscillations between positive and negative values. This emphasizes more complex and fluctuating spin interactions in larger systems. The amplitude of these oscillations decreases at longer distances, but the correlations do not vanish entirely, suggesting some degree of long-range order.
The corresponding table \ref{tab:spin_correlation_all_spins_with_dashes} further confirms these observations. Nearest neighbor correlations (distance $1$) show weak anti-correlations in the $4$-spin and $16$-spin systems, while the $8$-spin system displays slight positive correlations. In neighbors next-nearest (distance $2$), all systems show small negative correlations, indicating weak anti-alignment. For intermediate distances ($3$ to $7$), the $8$-spin system shows weak positive correlations, while the $16$-spin system oscillates among positive and negative correlations. At longer distances ($8$ to $15$) in the $16$ spin system, the correlations oscillate around zero and weaken further, reflecting the finite nature of the system and the limitation of the long-range order. Finally, they decrease with distance in all systems, with the $16$-spin system displaying more intricate oscillations due to its size. Spin interactions are predominantly local, with weak and diminishing correlations at larger distances.

\begin{figure}[h]
    \centering
    \includegraphics[width=0.5\textwidth]{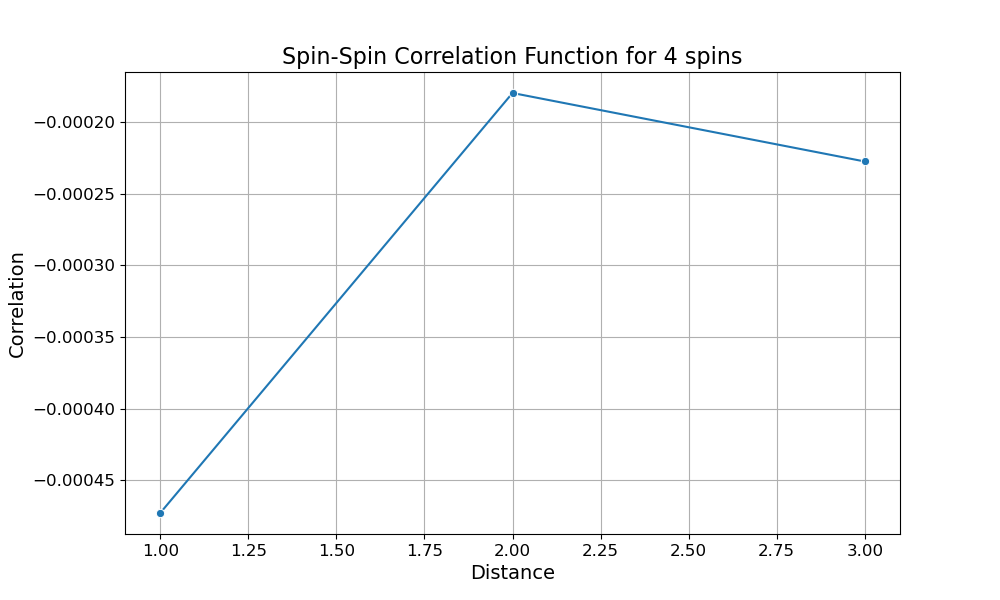}
    \caption{Spin-Spin Correlation Function for 4 Spins.}
    \label{fig:spin_spin_correlation_4}
\end{figure}

\begin{figure}[h]
    \centering
    \includegraphics[width=0.5\textwidth]{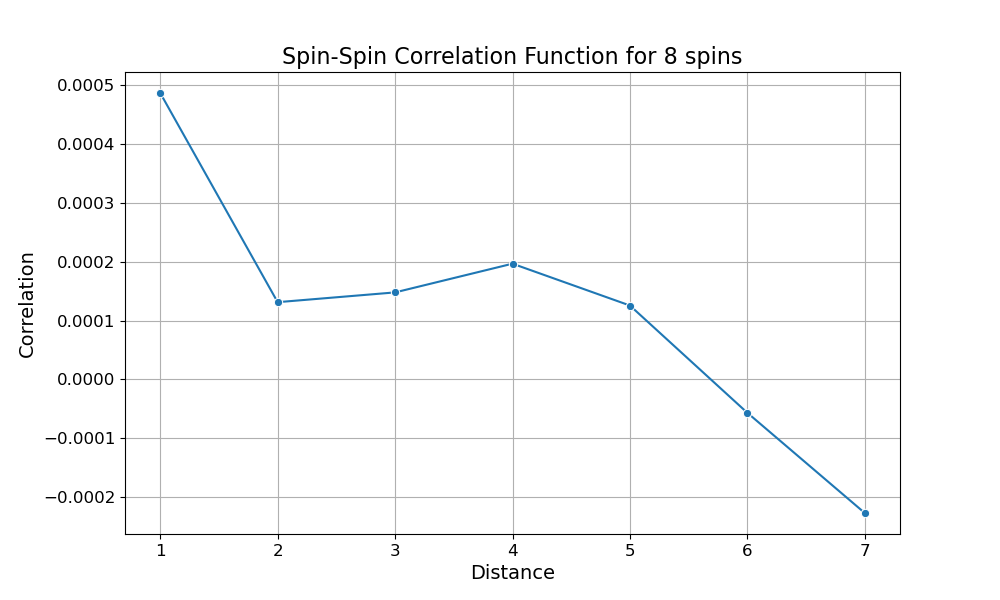}
    \caption{Spin-Spin Correlation Function for 8 Spins.}
    \label{fig:spin_spin_correlation_8}
\end{figure}

\begin{figure}[h]
    \centering
    \includegraphics[width=0.5\textwidth]{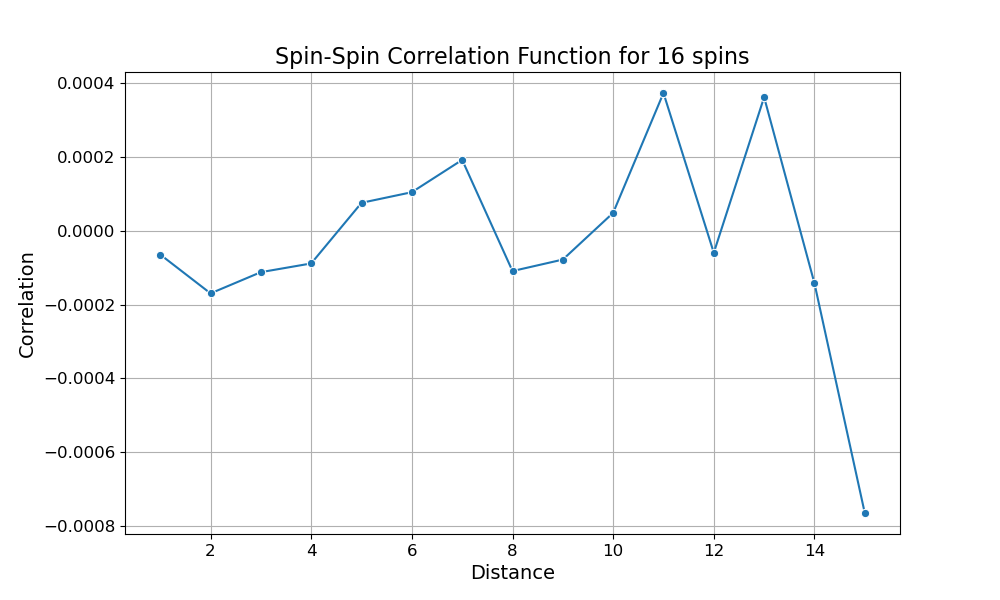}
    \caption{Spin-Spin Correlation Function for 16 Spins.}
    \label{fig:spin_spin_correlation_16}
\end{figure}

\begin{table}[h!]
\centering
\resizebox{\columnwidth}{!}{%
\begin{tabular}{|c|c|c|c|}
\hline
\textbf{Distance} & \textbf{4 Spins Correlation} & \textbf{8 Spins Correlation} & \textbf{16 Spins Correlation} \\ \hline
1 & -0.000473 & 0.000487 & -0.000065 \\ \hline
2 & -0.000179 & 0.000131 & -0.000170 \\ \hline
3 & -0.000227 & 0.000148 & -0.000112 \\ \hline
4 & - & 0.000197 & -0.000089 \\ \hline
5 & - & 0.000126 & 0.000076 \\ \hline
6 & - & 0.000217 & -0.000014 \\ \hline
7 & - & 0.000104 & 0.000061 \\ \hline
8 & - & - & -0.000095 \\ \hline
9 & - & - & 0.000039 \\ \hline
10 & - & - & 0.000054 \\ \hline
11 & - & - & -0.000031 \\ \hline
12 & - & - & -0.000003 \\ \hline
13 & - & - & -0.000027 \\ \hline
14 & - & - & 0.000016 \\ \hline
15 & - & - & 0.000001 \\ \hline
\end{tabular}%
}
\caption{Spin-Spin Correlation for 4, 8, and 16 Spins}
\label{tab:spin_correlation_all_spins_with_dashes}
\end{table}
\newpage

\item{Fluctuation distributions}

The distributions of the derivative of energy with respect to the external magnetic field, $\partial E / \partial H$, are shown in Figure \ref{fig:distribution_dE_dH}. The wider spread for the $16$-spin system compared to the $4$- and $8$-spin systems points to that larger systems show extra strong energy fluctuations in reply to changes in the external magnetic field. This indicates an increased sensitivity to external conditions in larger systems, which is characteristic of systems approaching a quantum phase transition.

\begin{figure}[h]
    \centering
    \includegraphics[width=0.5\textwidth]{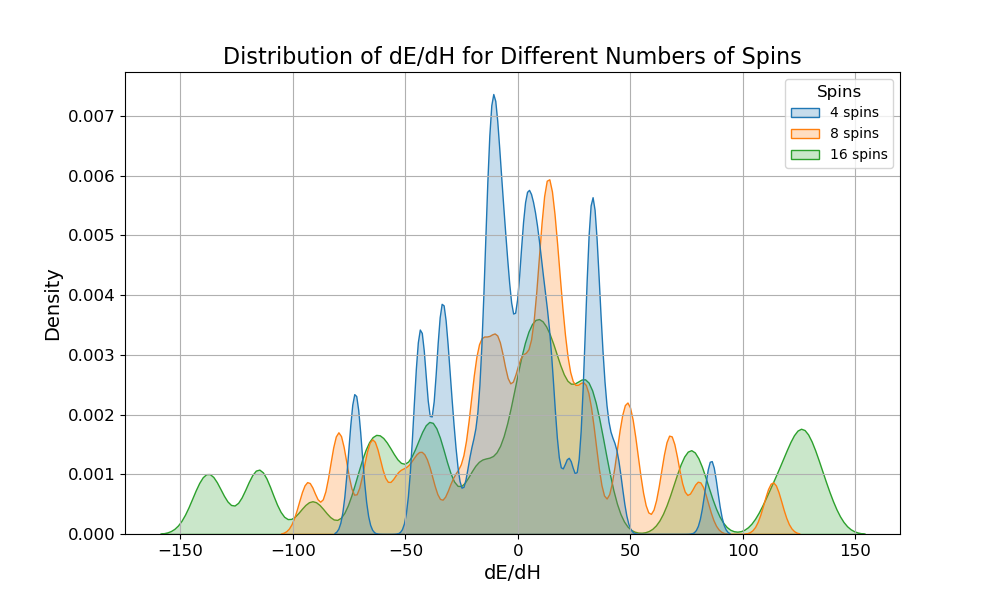}
    \caption{Distribution of $\partial E / \partial H$ for Different Numbers of Spins.}
    \label{fig:distribution_dE_dH}
\end{figure}
\newpage

\item{Energy Fluctuations with External Magnetic Field}

Figures \ref{fig:dE_dH_4}, \ref{fig:dE_dH_8}, and \ref{fig:dE_dH_16} disclose the derivative of energy with respect to the external magnetic field for systems with $4$, $8$, and $16$ spins. As the size of the system increases, the fluctuations in $\partial E / \partial H$ become more noticeable, with the $16$-spin system exhibiting sharp peaks and valleys. These results indicate that larger systems are high susceptible to energy fluctuations as a result of quantum effects, which may be indicative of increased instability or phase transitions.

\begin{figure}[h]
    \centering
    \includegraphics[width=0.5\textwidth]{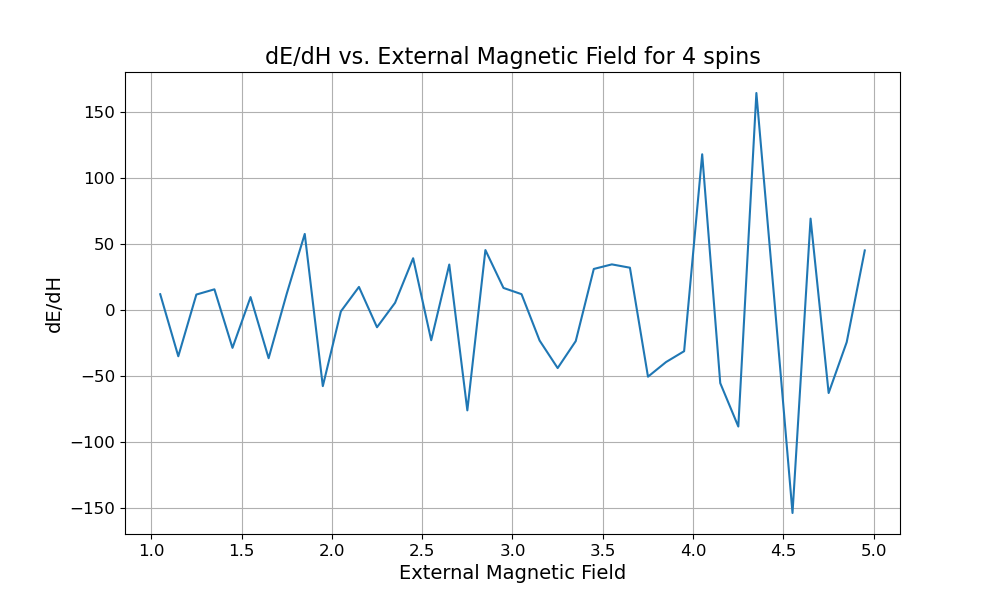}
    \caption{$\partial E / \partial H$ vs. External Magnetic Field for 4 Spins.}
    \label{fig:dE_dH_4}
\end{figure}

\begin{figure}[h]
    \centering
    \includegraphics[width=0.5\textwidth]{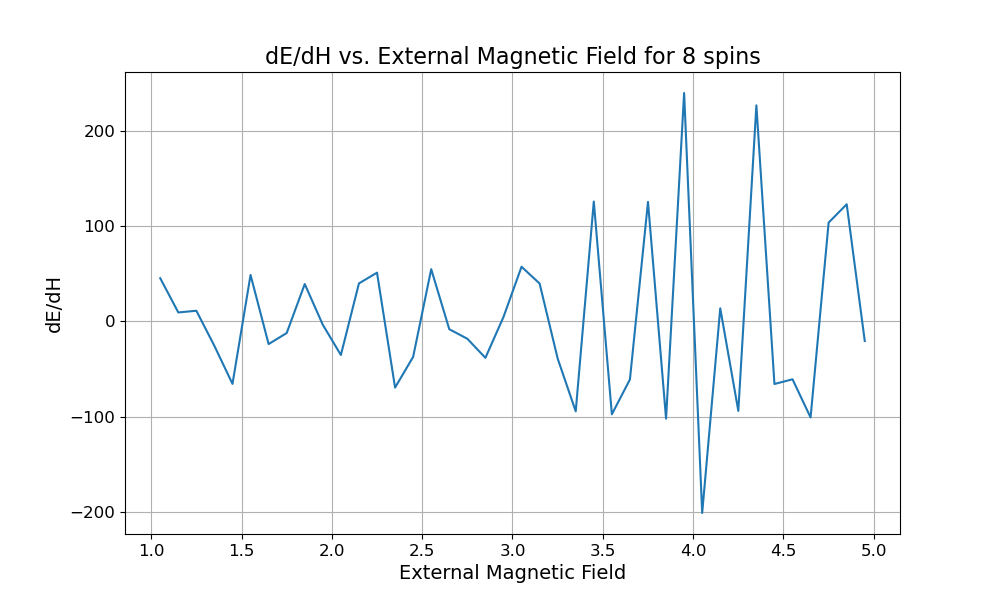}
    \caption{$\partial E / \partial H$ vs. External Magnetic Field for 8 Spins.}
    \label{fig:dE_dH_8}
\end{figure}

\begin{figure}[h]
    \centering
    \includegraphics[width=0.5\textwidth]{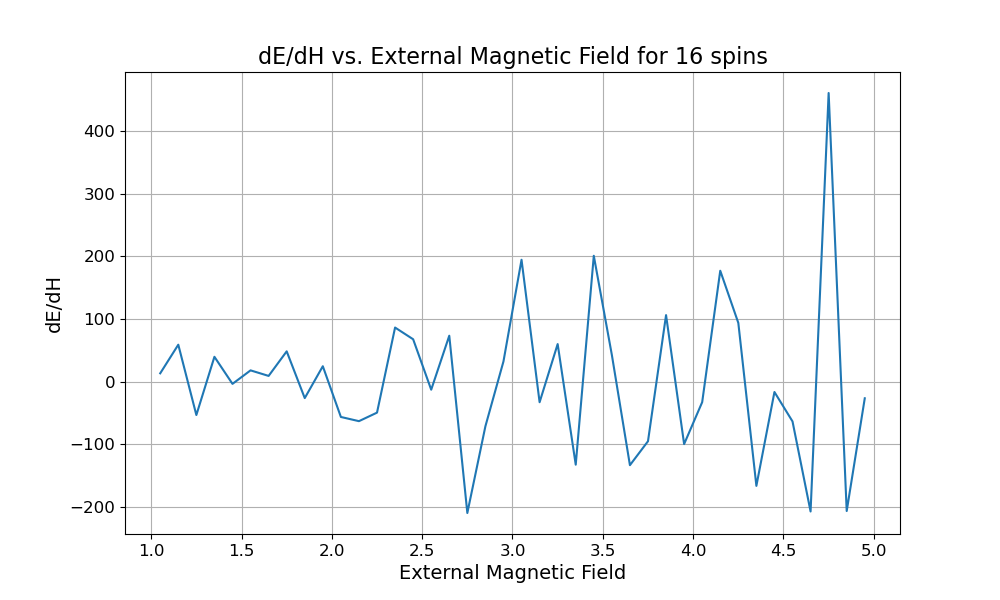}
    \caption{$\partial E / \partial H$ vs. External Magnetic Field for 16 Spins.}
    \label{fig:dE_dH_16}
\end{figure}

Furthermore, the second derivative of energy with respect to the external magnetic field, $d^2E/dH^2$, was computed and plotted for the spin systems, as shown in Figures \ref{fig:d2E_dH2_4}, \ref{fig:d2E_dH2_8}, and \ref{fig:d2E_dH2_16}, respectively. For the $4$-spin system (Figure \ref{fig:d2E_dH2_4}), the second derivative fluctuates with both positive and negative values, indicating non-monotonic behavior of the energy in response to changes in the external magnetic field. These fluctuations indicate that the energy reply to the magnetic field is not smooth, and there are small energy shifts at many field strengths. In the $8$-spin system (Plot \ref{fig:d2E_dH2_8}), the amplitude of the fluctuations increases. The plot shows prompt changes in the second derivative, with abrupt positive and negative peaks indicating evident shifts in the system’s energy when subjected to external magnetic field variations. This points to an increased sensitivity to changes in the magnetic field, possibly due to complex spin interactions in this system size. For the $16$-spin system (Figure \ref{fig:d2E_dH2_16}), the fluctuations become even more extreme, with peaks reaching higher positive and negative values than in smaller systems. This behavior points to that the latter exhibits significant energy shifts in response to changes in the external magnetic field. The large oscillations indicate complex interactions, possibly hinting at the onset of phase transitions or critical behavior at certain field strengths. In the plot \ref{fig:d2E_dH2_distribution}, we have the distribution of the second derivative of energy, $\frac{d^2 E}{d H^2}$, for our spin systems that also shows that larger systems are approaching critical behavior, with enhanced energy reply to field variations. Therefore, in summary, as the system size increases, the second derivative of energy with respect to the external magnetic field displays more pronounced oscillations. This suggests that larger systems exhibit difficult energy landscapes and greater sensitivity to external magnetic perturbations.

\begin{figure}[h]
    \centering
    \includegraphics[width=0.5\textwidth]{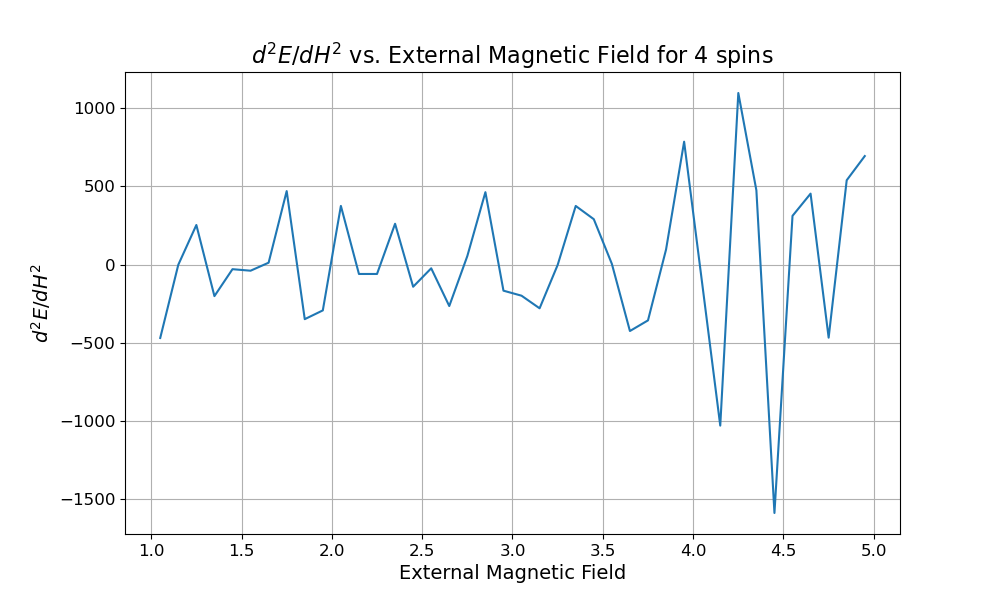}
    \caption{Second Derivative of Energy with Respect to the External Magnetic Field for 4 Spins.}
    \label{fig:d2E_dH2_4}
\end{figure}

\begin{figure}[h]
    \centering
    \includegraphics[width=0.5\textwidth]{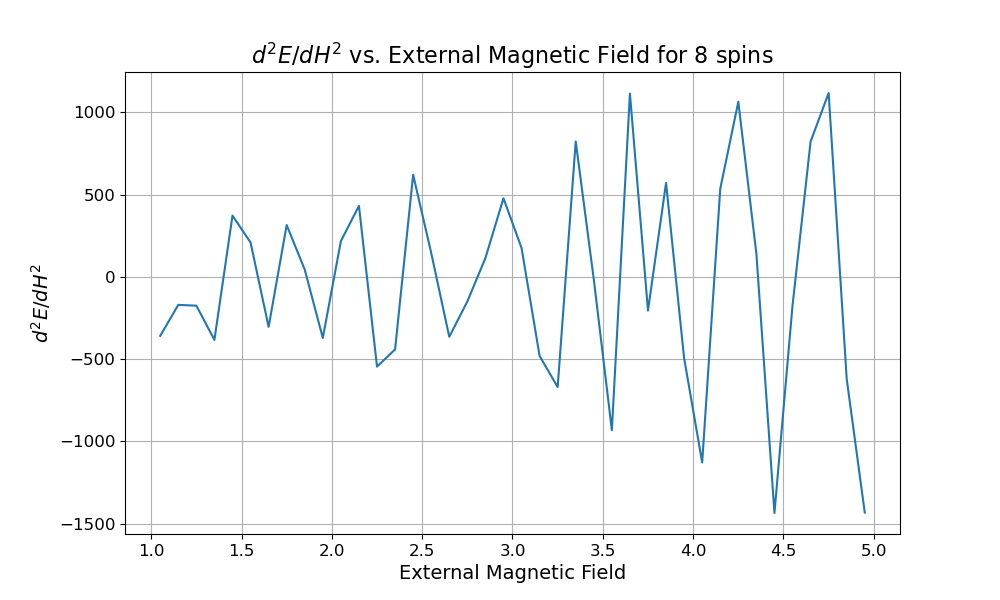}
    \caption{Second Derivative of Energy with Respect to the External Magnetic Field for 8 Spins.}
    \label{fig:d2E_dH2_8}
\end{figure}

\begin{figure}[h]
    \centering
    \includegraphics[width=0.5\textwidth]{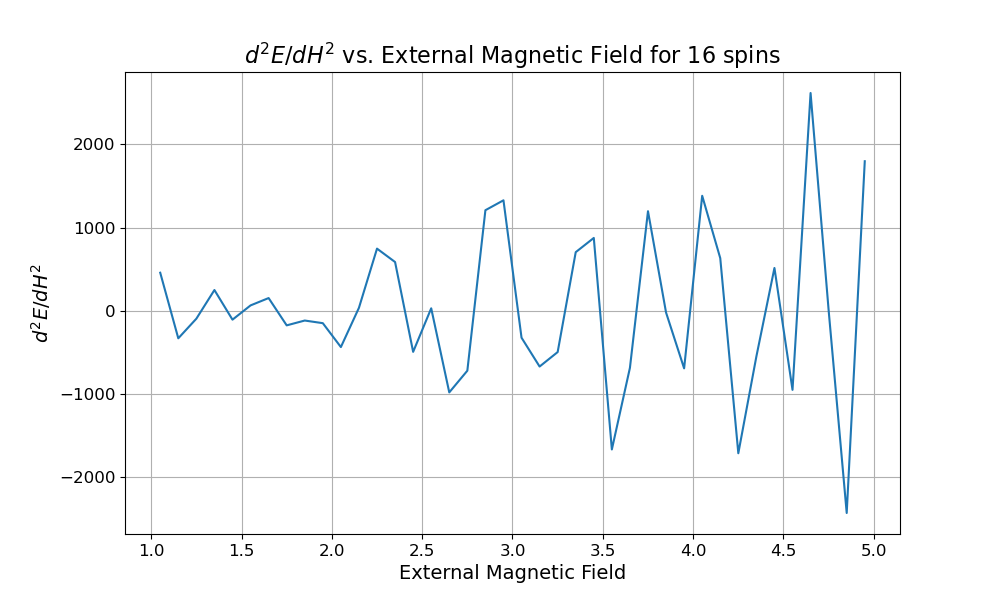}
    \caption{Second Derivative of Energy with Respect to the External Magnetic Field for 16 Spins.}
    \label{fig:d2E_dH2_16}
\end{figure}

\begin{figure}[h]
    \centering
    \includegraphics[width=0.5\textwidth]{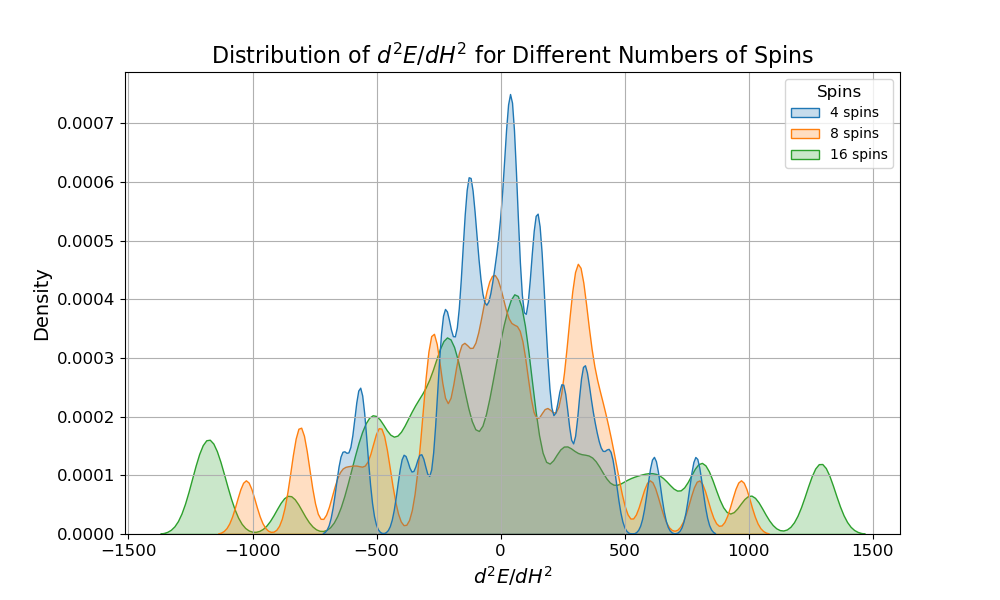}
    \caption{Distribution of $\partial^{2} E / \partial H^{2}$ for Different Numbers of Spins.}
    \label{fig:d2E_dH2_distribution}
\end{figure}
\end{itemize}

\newpage

\section{Conclusion} \label{sec:conclusion}

In this work, we have conducted an analysis of two-dimensional quantum spin systems using the quantum Ising model at absolute zero temperature. By simulating systems with $4$, $8$, and $16$ spins arranged in square lattices, we examined how fundamental physical properties---energy, magnetization, and entanglement entropy---evolve under varying external transverse magnetic fields.

Our simulations, performed using the Quantum Toolbox in Python (QuTiP), revealed many fundamental aspects.

\begin{itemize}
    \item \textbf{Energy Fluctuations:} As the size of the system increased, we observed significant fluctuations in energy in reply to the external magnetic field. The $16$-spin system exhibited much higher energy fluctuations compared to the other ones, indicating that larger systems become more sensitive to external perturbations. This behavior suggests the onset of quantum phase transitions, where larger systems are more susceptible to fluctuations.

    \item \textbf{Magnetization Behavior:} All systems showed an increase in magnetization with the growth of the external magnetic field. However, the magnetization response became more linear in the $8$-spin system, while the other systems displayed non-linear behaviors, possibly due to more localized spin interactions.

    \item \textbf{Entanglement Entropy:} The distribution of the entropy of entanglement varied with the size of the system. The $4$-spin system displayed a wide variability with several peaks, indicating a diversity of entanglement states. As the number of spins got bigger, the distribution turned out to be centralized around lower entropy values, reflecting increased coherence in larger systems.

    \item \textbf{Spin-Spin Correlations:} Our analysis of the spin-spin correlation functions revealed that the interactions are predominantly local, with the correlation strength decreasing as the distance between spins grew up. Larger systems exhibited more complex and fluctuating correlations, with the $16$-spin system showing oscillations among positive and negative values, indicating intricate spin interactions.

    \item \textbf{Energy Derivatives:} The first and second derivatives of energy with respect to the external magnetic field became evident with increasing system size. Larger systems showed sharper peaks and valleys, indicating increased sensitivity to external conditions and suggesting the possibility of critical behavior or phase transitions at certain field strengths.
\end{itemize}

These findings demonstrate that the behavior of quantum spin systems turns out to be increasingly complex as the system size grows. The results also underscore the potential for using our simulations to gain deeper intuition into the quantum behavior of spin systems.
Future work will focus on using these datasets for machine learning applications, and we hope to develop models that can accurately predict critical points and better understand the underlying mechanisms driving quantum phase transitions; any progress in this direction will be reported soon \cite{Terin2025}.

\section*{Acknowledgments}

 R. C. Terin acknowledges the Basque Government (projects $214023$FMAO, $214021$ELCN and IT$1504-22$).

\bibliographystyle{IEEEtran}    
\bibliography{phase_transition}

\end{document}